# Salt concentration dependent nucleation rates in low-metastability colloidal charged sphere melts containing small amounts of doublets

J. Schwarz[1], P. Leiderer[2], T. Palberg[1]*

[1]Institute of Physics, Johannes Gutenberg University, Mainz, Germany

[2]Fachbereicht Physik, University of Konstanz, Konstanz, Germany
*Corresponding author: palberg@uni-mainz.de

**Abstract**

We determined bulk crystal nucleation rates in aqueous suspensions of charged spheres at low metastability. Experiments were performed in dependence on electrolyte concentration and for two different particle number densities. The time-dependent nucleation rate shows a pronounced initial peak, while post-solidification crystal size distributions are skewed towards larger crystallite sizes. At each concentration, the nucleation rate density initially drops exponentially with increasing salt concentration. The full data set, however, shows an unexpected scaling of the nucleation rate densities with metastability times the number density of particles. Parameterization of our results in terms of Classical Nucleation Theory reveals unusually low interfacial free energies of the nucleus surfaces and nucleation barriers well below the thermal energy. We tentatively attribute our observations to the presence of doublets introduced by the employed conditioning technique. We discuss the conditions under which such small seeds may induce nucleation.

# I. INTRODUCTION

The phase behaviour and crystallization kinetics of colloidal suspensions with spherical particles have been studied extensively [1, 2, 3, 4, 5, 6]. They are recognized as valuable models for comparison to theory and simulation [7, 8] and for comparison to selected atomic systems like metals [2, 9]. For colloidal suspensions, the pair interactions, respectively the thermodynamic state and the properties of the suspension can be conveniently tuned at constant ambient temperature. Hard-sphere (HS) like systems are controlled via the HS packing fraction, while in depletion attractive HS, the strength and range of attraction can be varied through the size and the concentration of depletants [10]. For charged spheres (CS, e.g. negatively charged polystyrene particles suspended in water), the screened electrostatic repulsion can be precisely tuned *via* the effective particle charge, $Z_{eff}$, the concentration of screening electrolyte, $c_s$, and the number density of particles, $n$ [11]. Their variation then controls phase behaviour [12, 13, 14] and crystallization kinetics [15], *via* the ratio of pair interaction energy to thermal energy [16]. However, due to the presence of the solvent acting as heat-bath, colloidal crystallization proceeds at constant temperature. However, colloidal crystals may be shear molten mechanically [17]. Further, typical length and time scales regarding suspension structure and dynamics as well as the nucleation, growth and coarsening kinetics are conveniently accessible by suitable optical methods [18]. Finally, tractability of interaction potentials allows for theoretical modelling and simulation. E.g., by combining confocal microscopy with advanced computational concepts, valuable insight was gained into melt pre-structuring, nucleus structure and the role of collective rearrangements [19, 20, 21, 22, 23, 24, 25, 26]

In general, but with some important exceptions [19, 27, 28], colloidal solidification kinetics involving homogeneous bulk nucleation can be parameterized by simple classical models [4, 29, 30]. In classical nucleation theory (CNT), nucleation is considered as a thermally activated process [7, 31, 32]. It is followed by growth, which in turn is well described as diffusion-limited (HS) or reaction-controlled (CS) within a Wilson-Frenkel approach [33, 34]. In addition, extensions of CNT to heterogeneous nucleation at extended surfaces and large seed particles are available [7, 8, 35]. At elevated number density, CS systems of low polydispersity and their respective mixtures have been addressed in a wealth of studies [29, 36, 37, 38]. Research continues with focus on the systematic influence of polydispersity [30], crystallization from the glassy state [39], and heterogeneous nucleation [40, 41]. Rather few experiments were performed on CS at low metastabilities. Growth studies have been reported [15,

42, 43, 44, 45, 46, 47], but studies on bulk nucleation kinetics are still very rare [48, 49, 50]. The present paper aims at filling this gap.

Our study explores the solidification kinetics of CS in the region at low metastability including the fluid-crystal coexistence region. It originally intended to study the salt concentration dependence of the homogeneous nucleation rate densities, $J$, at constant number density. The investigated system shows a low-lying freezing transition in the deionized state at $n_F = 3.06 \mu m^{-3}$. We prepared two systems at number densities of $n = 5.4 \mu m^{-3}$ and $n = 9.5 \mu m^{-3}$. Using a peristaltically driven conditioning circuit under inert gas atmosphere, we varied the concentration of added electrolyte under conductometric control in small steps. To work at strictly constant number density, we omitted the usually implemented filter. This, however, introduced a significant amount of particle doublets. Sample composition analysis later yielded an upper bound for the doublet fraction of $f_D = <n_{doublet} / n> \leq 10^{-3}$.

Conditioned samples were shear molten and their recrystallization followed by optical methods. We used Bragg-video-microscopy as well as post solidification image analysis. Nucleation experiments then yielded the salt concentration dependence of nucleation rate densities in the presence of doublets. Additional growth experiments allowed calibration of the metastability in terms of the chemical potential difference, $\Delta \mu$, between melt and crystal as a function of number density and salt concentration. Together this facilitated the desired quantitative characterization of the nucleation kinetics way into the coexistence region and under a finely spaced variation of the metastability. However, the nucleation kinetics observed here differ considerably from those observed at large meta-stabilities in previous experiments.

To be more specific, we find peaked nucleation rates down to lowest meta-stabilities, and crystallite size distributions are pronouncedly skewed towards larger crystals. For each $n$, the observed $J$ increased exponentially with increasing $\Delta \mu$, but a three orders of magnitude discontinuity was observed between data for the two $n$. Moreover, subjecting our data to parameterization by CNT, we extracted very small activation barriers, which would correspond to extremely small interfacial free energies (IFE). Such a behaviour is not compatible with expectations of CNT for homogeneous bulk nucleation from an isotropic, homogeneous and ergodic melt. Tentatively assuming the observed doublets to act as small seeds, we find the seeding efficiency and the heterogeneous nucleation rate to increase exponentially with $n\Delta\mu$. These unexpected findings are beyond the scope of CNT in its extensions to heterogeneous nucleation and call for alternative approaches.

## II. SAMPLE CHARACTERIZATION AND CONDITIONING

We studied commercial Polystyrene latex particles stabilized by N = 1200 strongly acidic surface groups, as determined from titration (PS109*; Seradyn lot #2011M9R). These particles had been extensively characterized and used as model systems in a variety of previous studies [45, 51, 52, 53, 54, 55, 56]. The *nominal* diameter given by the manufacturer is 109nm. The nominal size dispersity is rather small with the ratio of mean and standard deviation of $s_a/<a>$ = 0.02. We found a diameter of $2a_H$ = (110.1±2)nm from dynamic light scattering, and $2a_{AFM}$ = (106±15)nm from force microscopy on particles deposited on oppositely charged substrates. For our calculations below we used the scattering radius $a = a_{PQ}$ = (51±1)nm as obtained from form factor measurements. In suspension, the effective charge numbers are smaller than the group number N. We determined the number of freely moving protons in the electrostatic double layer from conductivity measurements on deionized samples in dependence on particle number density $n$. We employed Hessinger's model of independent ion migration [57, 58] to obtain $Z_\sigma$ = 459$e^-$ (c.f. Fig. A1(a) in Appendix A). This number is close to the expectation from mean field cell-model calculations and used in the determinations of the electrolyte concentration. The interaction strength is calculated using the effective charge from elasticity measurements on crystallized samples in dependence on $n$ [59], which amounts to $Z_G$ = 321$e^-$ (c.f. Fig. A1(b) in Appendix A).

After pre-conditioning, we employed a closed preparation cycle for sample conditioning, in which the suspension is peristaltically driven through a gas-tight tubing system connecting different circuit components (for details, see Appendix A). This systems allows fast deionization and reproducible adjustment of $c_s$ in the μmolar range [60]. It further facilitates integration of several measurements to be performed under identical preparation conditions. A suitable amount of stock suspension is filled under filtering into the circuit and further diluted with distilled water to the desired concentration. As this results in some particle loss, we can check the final particle density by static light scattering. A flow-through ion-exchange chamber filled with mixed bed ion exchange resin is used for deionization. The sample conductivity is monitored to equilibrate under continued cycling. In the deionised state, residual electrolyte concentrations are dominated by the particle counter-ion contribution and the self-dissociation of water. NaCl solution is then added under by-passing the ion-exchange cell, and conductivity readings are converted to concentrations of added electrolyte, $c_s$, using Hessinger's model with $n$ and $Z_\sigma$ as input [57]. The circuit also contains the actual measuring cell mounted to the stage of an inverted microscope [Leica IRB, Leitz, Wetzlar Germany]. Circulation leaves the sample in a well homogenized, metastable shear molten state. Stopping the circulation defines the start of the solidification experiments at t = 0.

By default, the circuit is equipped with a continuous filtering to remove any contaminations like coagulate or resin splinters either residual or formed during cycling. Potential particle loss then necessitates online monitoring of the particle density. This version was employed for the determination of the phase diagram. For the nucleation and growth measurements, we however omitted the filter, to keep the number density strictly constant. We initially regarded contamination as negligible for our experiments. However, after obtaining unexpected results, we addressed this issue explicitly, as contaminations may potentially act as nucleation seeds.

Particles are too small for size discrimination by optical microscopy. We checked for the presence of contaminations using particle adsorption followed by scanning force microscopy. We cut 4x70mm pieces from standard glass slides spin-coated with Polyvinylpyridine, which were dipped into the suspension. The negatively charged particles, but also doublets, dust and ion exchange debris readily adsorb onto the coated slides. Particles cover the substrate fairly homogeneously at a typical mutual distance but with no long range order. The slides were scanned with an Atomic Force Microscope (AFM, DI, Germany; tapping mode, scan frequency 2 Hz, scan range 3.1x3.1µm$^2$). Fig. 1 shows three examples, with the $z$-scale colour coded (brown, yellow and white corresponding to heights of z = 0nm, z ≈ 100nm and z ≥ 200nm, respectively). For technical reasons, the particles shapes appear to be smeared out to the left. Fig. 1a shows a typical slide with no seed candidates. In fact, in most scanned areas, none or just one candidate particle was observed. Fig. 1b shows a larger object in the lower middle. Such objects are easily identified by their increased height and diameter but very rare. Fig. 1c shows an object of approximately doubled height and minimally increased diameter. This is identified as doublet, deposited in an upright position.

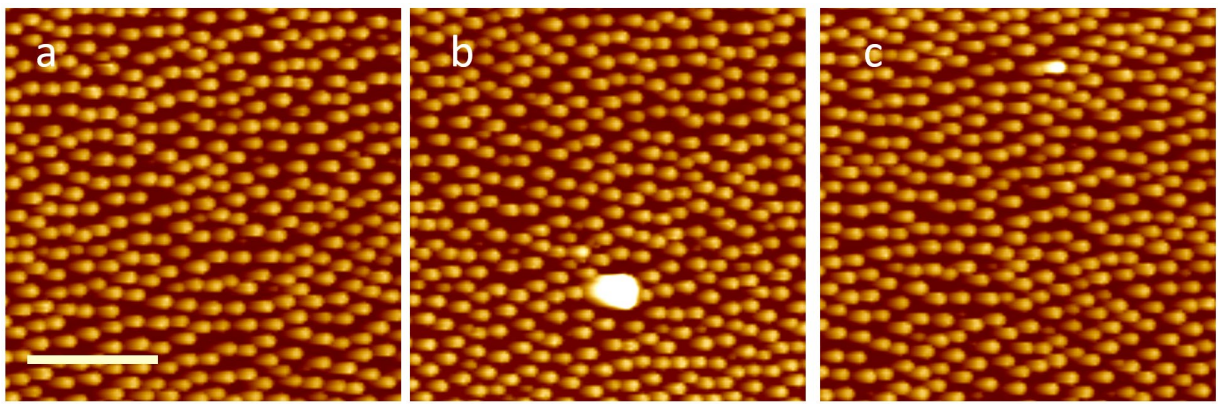

Fig. 1: AFM images of PS109* particles deposited on Polyvinylpyridine-coated glass slides upon dipping them into the suspension. Scale bar 1µm. The colour-coded *z*-range covers 200nm from the substrate (*z* = 0nm). a) Typical scan with no impurities. b) larger impurity in the lower middle. c) Doublet in the upper middle.

Seed fractions were estimated for test slides prepared in suspensions conditioned for different stretches of time. Accuracy then depends on statistics. A total of 150 scans of 3.1x3.1µm$^2$ (catching some 530 particles in each) at different positions along the slides did not show variations in particle coverage nor preferred deposition regions for larger objects. For a sample conditioned for half an hour, we found 15 doublets within 8 10$^4$ particles yielding the number fraction of doublets of $f_D$ = <N$_{impurity}$ / N$_{particle}$> ≈ (2±1) 10$^{-4}$. After 6h of cycling we found 81±10 deposited doublets (average over eight slides). This yielded $f_D$ ≈ (1±0.1)10$^{-3}$. We detected few larger objects ($f_L$ ≤ 10$^{-5}$). Interestingly, no triplets were observed. Integrating a 0.4µm pore size filter and cycling for one hour reduced the impurity fraction considerably to $f_D$ ≤ 10$^{-5}$ (no larger objects). At the same time, it reduced the number density by some 8%. We take the values obtained for 6h of cycling as an upper limit. The upper bound for the seed number density then is $n_{seed}$ ≤ 5.4 10$^{15}$m$^{-3}$ and $n_{seed}$ ≤ 9.5 10$^{15}$ m$^{-3}$ in the two crystallization experiments, respectively.

## III. RESULTS AND DISCUSSION

### A. Phase behaviour and growth measurements

The phase diagram of PS109* is shown in Fig. 2(a). At low *n* and large $c_s$ the suspension is in a fluid state, at larger *n* and lower $c_s$, samples crystallize in a bcc structure. The coexistence region widens with increasing *n* and $c_s$. The behaviour is fully consistent with observations on other particle species. As compared to the phase diagram for the less charged twin particle species (PS109; Seradyn lot #2010M9R; [15]), the crystalline region extends to somewhat higher salt concentrations, but otherwise is very similar. The location of the melting line is in good agreement with expectations from simulation [16]. In Fig. 2(a), we further display the range of salt concentration covered in our three kinetic experiments (hatched bars) and the location of the observed freezing points (endpoints of the dashed arrows).

Samples prepared in the meta-stable fluid state readily recrystallize, either *via* growth after heterogeneous nucleation at the cell wall, or *via* bulk nucleation followed by growth. Wall crystal growth was monitored using Bragg microscopy under external white light illumination in a cell of rectangular cross section (10x1mm², Lightpath Optical, UK) [15]. Figure 2(b) compares the data for two independent runs and demonstrates excellent reproducibility. In the coexistence region (red hatched bar), the recorded velocities slow as a function of time as the extension of the wall crystal approaches its equilibrium value. In this case, we noted the initial growth velocities. In the crystal region of the phase diagram, the wall crystal fronts advance linearly in time up to intersection with either the opposing crystal or with bulk nucleated crystals. Here we determined the average velocity by fitting a linear function.

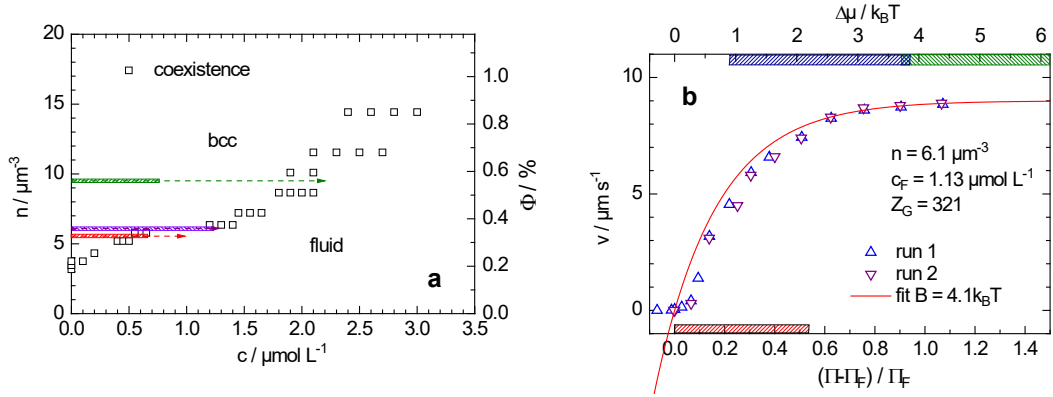

FIG. 2: Phase behaviour and growth measurements. (a) Phase diagram of PS109* in dependence on concentration of added electrolyte and number density. Open squares denote samples, for which coexisting phases were observed. Arrows and hatched areas denote the experimental $n$, the observed freezing points and the range of added electrolyte in the solidification kinetic measurements, respectively. Colours discriminate between the three experiments; red: nucleation experiments at $n = 5.4\mu m^{-3}$; green: nucleation experiments at $n = 9.5\mu m^{-3}$; violet: growth experiments at $n = 6.1\mu m^{-3}$. (b) Growth velocities of wall nucleated crystals as a function of the rescaled energy density $\Pi^*$ as calculated from the experimental parameters and the location of freezing. A WF-growth law fit to the data returns a calibration coefficient of $B = (4.1\pm 0.3)k_B T$. Blue and green hatched bars at the top show the matching ranges of metastability covered in the nucleation experiments at $n = 5.4\mu m^{-3}$ and $n = 9.5\mu m^{-3}$, respectively. The red hatched bar indicates the extension of the coexistence region at $n = 5.4\mu m^{-3}$.

We assessed the metastability of the melt state following the procedure suggested originally by Aastuen [42], and later improved by Würth [15] and Palberg [18]. We exploit the fact that wall crystal growth follows a Wilson-Frenkel (WF) law for reaction-controlled growth in our system: $v = v_\infty (1-\exp(-\Delta\mu/k_B T)$, where $k_B T$ denotes the thermal energy. We assume the chem-ical potential difference, $\Delta\mu$, between the melt state and the emerging crystal to be proportional to a reduced energy difference, $\Pi^* = (\Pi-\Pi_F)/\Pi_F$, between the melt and the fluid at freezing (F), i.e. $\Delta\mu = B\Pi^*$. Here, $\Pi = \alpha n V(d_{NN})$ is an energy density calculated from the pair energy of interaction at the nearest neighbour distance, $V(d_{NN})$. The latter is modelled as screened electrostatic interaction between charged hard spheres and depends on $Z_G$, $n$ and $c_s$ (see Eqns.(A2) and (A3) in Appendix A). $\alpha$ is an effective coordination number. The parameter $B$ is then derived from a fit to data obtained in dependence on any of the three control parameters entering $V(d_{NN})$. In the present growth experiments, $n$ and $Z_G$ are kept constant, and only $c_s$ is varied. Figure 2(b) displays the growth velocities in dependence on $\Pi^*$ calculated with $c_F = (1.13\pm0.06)\mu molL^{-1}$. With increasing metastability, v first increases then saturates at a well-defined maximum value of $v_\infty = 9.0$ $\mu ms^{-1}$.

The figure further shows a least squares fit of a WF-law to the data describing the data very well. As expected, we find a small systematic deviation across the coexistence region related to the density difference between melt and crystal. The deviation increases as freezing is approached but still remains small for the low polydispersity system of PnBAPS109*. From the fit, we extract a coefficient $B = 4.1\pm0.3k_BT$ allowing to calibrate $\Delta\mu$ (upper scale). Note the slight overlap of the metastability ranges covered by the salt concentration dependent nucleation experiments at $n = 5.4\mu m^{-3}$ and $n = 9.5\mu m^{-3}$ (indicated by the blue and green hatched bars at the top of Fig. 2(b)). This enables the discrimination of the effects of $n$ and $c_s$ in the nucleation experiments.

### B. Nucleation experiments

For the bulk nucleation experiments, we used cells with larger depth (5mm) to avoid interference with wall crystal growth. We placed them on the stage of a polarization microscope and observed the sample in transmission between crossed polarizers. This provides a colour contrast between melt and crystals. Only for the latter, the polarization direction of the transmitted light is turned due to dynamical Bragg scattering in dependence on crystal orientation [61]. We used a low magnification objective (Leitz, HC PL Fluotar 5x/0.15 POL) to cover a volume $V_0$ of 5mm radius and 5mm depth. Nucleation rates show a pronounced variation with both salt concentration and particle number density. At $n = 5.4\mu m^{-3}$, nucleation was slow enough to be directly followed. We performed a direct counting of nucleation events from the recorded videos for time intervals of $\Delta t = 2.5$s. In Fig. 3(a) we display the number of nucleation events as a function of time. The nucleation rate is obtained dividing N by the respective time interval. The rate density follows from dividing the rate by the known volume.

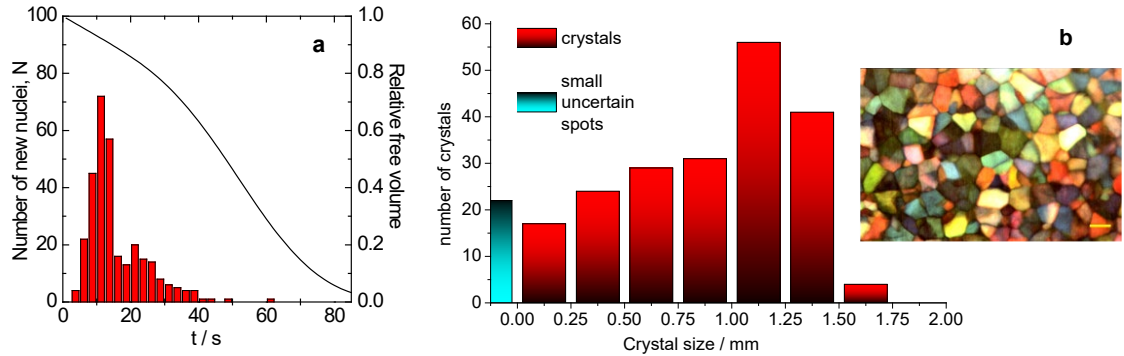

FIG. 3: Nucleation results. (a) Time dependent nucleation rates determined from direct video-microscopic observation. The solid line shows the evolution of the fraction of leftover melt volume (right scale). (b) Crystal size distribution for the post-solidification snapshot shown in the inset. Scale bar 1000µm, image size 8×12mm². Small regions with insufficient colour contrast are frequently observed at three-crystallite intersections. We here file them under "small uncertain spots".

After a short induction period the nucleation rate steeply rises to a narrow maximum but quickly settles back to a very low value. Figure 3(a) also displays the remaining free volume. The latter was calculated following Wette [48], i.e. using a suitable adaptation of Avrami's theoretical model [62, 63]. It accounts for the volume of newly formed and growing bulk crystals as well as for the wall crystal volume. For the former quantity, we used the directionally averaged bulk growth velocity of $1.05v_{110}$. For the latter we used the known crystal growth velocity in (110) direction, $v_{110}$, at this salt concentration (0.1 µmolL$^{-1}$, c.f Fig. 2(b)). The initial linear drop is due to wall crystal growth. The curve steepens as significant bulk crystal growth occurs, and flattens again due to crystallite intersection. Clearly, the nucleation rate drops long before any significant amounts of suspension have solidified. Most importantly, this peaked behaviour did not vanish at larger salt concentrations.

In Fig. 3(a), the initial rise in nucleation rate is fast and very short. For salt concentrations up to $c_s = 0.46$µmolL$^{-1}$, however, rates were large enough and statistical scatter low enough to assign a peak nucleation rate density using the percentage of leftover free volume and the observation volume. The peak nucleation rate densities displayed in the semi-logarithmic plots of Fig. 4(b) and 4(c) as open red circles show a roughly linear decrease with salt concentration and a roughly linear increase with metastability, respectively. In the latter plot, we also show the error bars arising for Δμ. These are due to the uncertainty of actual salt concentration in the calculation of Π* and the uncertainty of the calibration factor $B$ but mostly due to the systematic uncertainty in the exact location of the freezing salt concentration.

We further performed post solidification size analysis [48]. In Fig. 3(b), we illustrate this for the sample at $c_s = 0.1 \mu molL^{-1}$ together with a post-solidification snapshot of the analysed sample in the inset. Here, but also at larger salt concentrations, we observe size distributions, which are clearly skewed towards larger crystallite sizes. This skewing direction is fully consistent with our time-dependent observations of a strongly peaked nucleation. Moreover, the crystallite edges observed after complete solidification are all straight. According to the model of Johnson and Mehl [64], this is a direct consequence of the nucleation burst at small times. From the obtained sizing data we inferred the average crystallite density $\rho_c$ and plot it in Fig. 4(a) as a function of salt concentration. Using Avrami-theory, we then infer the effective steady state nucleation rate density $J$:

$$J^{ST}_{AVR} = 1.158 v \rho^{(4/3)} \qquad (1)$$

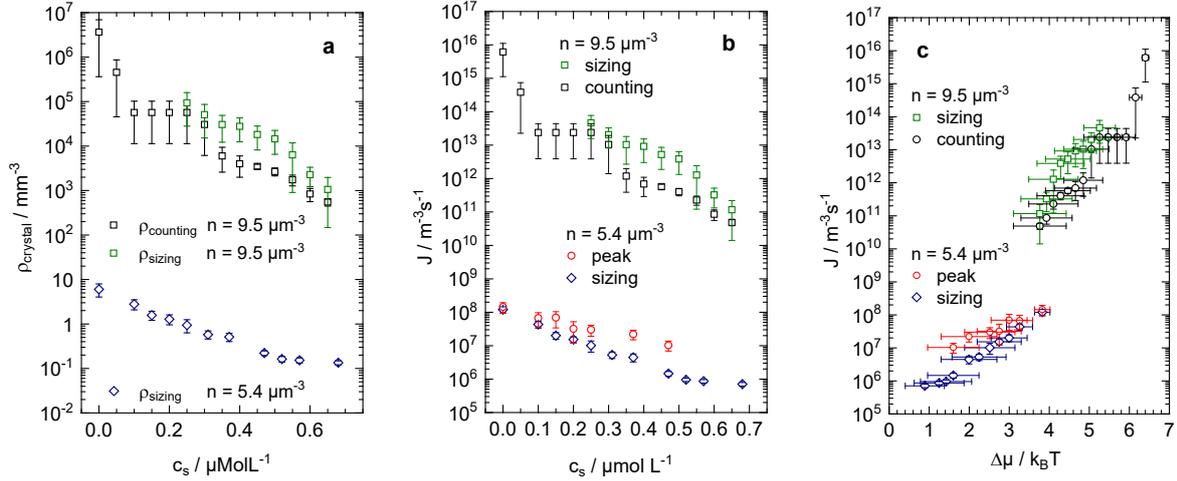

FIG. 4: Results of crystallization experiments. (a) Crystallite densities versus concentration of added salt. As indicated in the key, densities were derived from sizing at $n = 5.4$ μm$^{-3}$ as well as from sizing and counting at $n = 9.5$μm$^{-3}$. Vertical error bars denote the combined statistical and systematic uncertainty of $\rho_c$. The error in salt concentration is below symbol size. (b) Nucleation rate densities as a function of concentration of added salt for the two different number

densities and evaluation methods as indicated in the key. (c) Nucleation rate densities as a function of metastability for the two different number densities and evaluation methods as indicated in the key. Here, the horizontal error bars denote the systematic uncertainty of $\Delta\mu$ arising from the uncertainty in the determination of the freezing salt concentration combining with the uncertainty in actual salt concentration in the calculation of $\Pi^*$ and the uncertainty of the calibration factor $B$.

In Fig. 4(b) and (c), we display the obtained nucleation rate densities in dependence on salt concentration, respectively the metastability, as blue open diamonds. With increasing $c_s$, the rate densities initially decrease in an approximately linear fashion in this semi-logarithmic plot. However, for salt concentrations larger than $c_s \approx 0.5\mu molL^{-1}$ the densities level off at a value of $J \approx 8\ 10^5 m^{-3}s^{-1}$. At low salt concentrations, the rates inferred from sizing meet with the peak nucleation rates. This behaviour is also seen in the dependence on metastability in Fig. 4(c). Note, that for the sizing measurements in dependence on $\Delta\mu$, the increase of rate densities accelerates with increased metastability. Thus at larger metastability, it is steeper than for the peak rate densities.

We now turn to the second series of samples prepared at the larger number density of $n = 9.5\mu m^{-3}$. Here, nucleation was too fast to follow it directly. We therefore resorted to crystallite sizing and crystal counting after complete solidification to determine the average crystallite density $\rho_c$ and the effective steady state nucleation rate densities. As compared to the sizing at the lower number density, we now faced significantly systematic uncertainties due to the smallness of crystallite sizes. In fact, sizing became possible at all only for salt concentrations larger than $c_s \geq 2.5\mu molL^{-1}$. Further, sizing and counting sometimes yielded significantly different results, and the crystallite densities from sizing appear to be systematically larger. This is presumably due to the different averaging procedures acting on the skewed size distributions. In Fig. 4(a), we compare the derived crystallite densities to the data obtained at the lower particle concentration. In dependence on salt concentration $\rho_c$ again changes by several orders of magnitude but now on a significantly higher level. The corresponding nucleation rate densities are shown in Fig. 4(b). They now drop from $J \approx 5\ 10^{15}m^{-3}s^{-1}$ in the deionized samples to $J \approx 10^{11}m^{-3}s^{-1}$ at $c_s = 0.7\mu molL^{-1}$. For each number density the data set shows a roughly linear decrease in J with increasing salt concentration in this semi-logarithmic plot. Thus, the qualitative dependence on salt concentration appears to be retained. When the data are replotted versus the metastability in Fig. 4(c), the data from both approaches to $\rho_c$ overlap reasonably well within the combined experimental uncertainties. We obtain an exponential increase of $J$ with $\Delta\mu$ for each of the data sets. The dependence of $J$ on metastability is somewhat steeper at the larger number density. However, at $\Delta\mu = 4k_BT$, there is a discontinuity, and a jump of about three orders in magnitude to larger $J$ can be seen at the same metastability. This discontinuous evolution of $J$ as a function of $\Delta\mu$ is unexpected.

### C. Data analysis and comparison to previous results

The survey above shows marked differences to previous findings. The discontinuous behaviour of the nucleation rate density as a function of metastability has not been seen before in homogeneous bulk nucleation. Further, the observation of peaked bulk nucleation down to very low metastabilities is very unusual. In both metals and charged spheres, previous studies showed an extended plateau with steady state nucleation rates framed by a pronounced initial increase and final decrease. For metals, the final decrease is attributed to the release of the heat of fusion resulting in a decrease of metastability [65]. For colloids, it was attributed to a dilution of the melt occurring upon forming the slightly denser crystallites [4]. However, for both metals and charged colloids, the extension of the plateau was found to *shrink* with increasing metastability, leading to peaked nucleation only at large metastability [2, 48, 9]. Then, the non-steady nucleation could be successfully interpreted as quickly terminating burst, and a corresponding steady state nucleation rate density could be estimated [2, 71]. By contrast, the here observed peaked behaviour extends way into the coexistence region, while the initial rise covers too few points to assign a steady state nucleation rate $J^{ST}$ from its initial slope.

To quantify our observations further and to compare to previous findings obtained at moderate to large metastability, we next parameterize our results within CNT for homogeneous bulk nucleation. CNT was originally developed as macroscopic theory to describe the condensation of droplets and later adapted to also describe crystallization. [66, 67, 68, 69, 70, 71]. We are aware of the many assumptions made within and the controversial discussion of this approach [8]. Still, it successfully captures the crystallization kinetics of some systems including simple metals [72] and charged colloidal spheres [4]. It then allows deriving estimates for the key parameters of homogeneous nucleation [30]. In essence, CNT views homogeneous nucleation as an activated process involving one-by-one particle addition. An energy barrier separates the initial homogeneous, isotropic melt (thought to share the structural properties of an equilibrium fluid) and the final perfect crystal phase. It emerges from the competition between the free energy gain obtained in forming the crystal and free energy needed to create the sharp interface of a spherical cluster: $\Delta G = (4\pi/3)r^3 \, n\Delta\mu + 4\pi r^2 \gamma$, where $\gamma$ is the interfacial free energy. Further, all structural details of the interface and the involved phases are ignored. Under these boundary conditions, the steady state CNT homogeneous nucleation rate reads:

$$J^{ST} = k^+(n^*)ZN_{eq}(n^*) = J_0 \exp\left(-\frac{\Delta G^*}{k_B T}\right) = J_0 \exp\left(-\frac{16\pi}{3}\frac{\gamma^3}{(n\Delta\mu)^2 k_B T}\right) \quad (3),$$

where n* is the number of particles in the critical cluster, and $k^+(n^*)$ is the attachment rate to the critical cluster. $Z = (|\Delta\mu|/6k_B Tn^*)$ is the Zeldovich factor accounting for the shape and width of the barrier [7, 67, 68, 73]. $N_{eq}(n^*)$ is the equilibrium number of critical clusters replaced in the second step by the Boltzmann term involving the barrier height, $\Delta G^*$. $k^+(n^*)$ and $Z$ are combined to give the kinetic pre-factor, $J_0$, characterizing the maximum possible steady state nucleation rate density.

For colloid crystal nucleation from the melt, $J_0$ depends only weakly on $n$ and $\gamma$ [29]. It scales with the long-time self-diffusion coefficient, which is approximated *via* the dynamic freezing criterion [53, 54, 74]. Neglecting the small dependence on $\gamma$ of the pre-factor, $\gamma$ is then obtained as the cube root of the $\Delta\mu$-dependent slope from a plot of the natural logarithm of $J$ versus $1/(n\Delta\mu)^2$. We illustrate this approach in Fig. 5(a). The main part of the figure shows all data and their dependence on $1/(n\Delta\mu)^2$. The inset exemplarily displays the 3 point and 4 point linear fits to the data obtained from sizing at $n = 5.4\mu m^-$ (additional material can be found in Appendix D, Figs. A4, A5 and A6). We normalize $\gamma$ by the area taken by a particle in the nucleus surface and express it in units of thermal energy: $\sigma = \gamma d^2_{NN}/k_B T$. Results for this reduced interfacial free energy of the sample at $n = 5.4\mu m^{-3}$ are displayed in Fig. 5(b). The vertical error bars here derive from the standard errors of the fits in Fig. 5(a) at a confidence level of 0.95. At $\Delta\mu = 3k_B T$, $\gamma = 4.5 nJm^{-2}$, respectively $\sigma = 0.25 k_B T$. Both data sets overlap within the fitting errors, and $\sigma$ decreases roughly linearly with decreasing $\Delta\mu$. However, it approaches zero close to $\Delta\mu \approx 1k_B T$. The solid lines are fits of $\sigma = \sigma_0 + C_T\Delta\mu$ to the data. They return a Turnbull coefficient of $C_T = (0.092\pm0.004)$ ($C_T = (0.084\pm0.017)$) for the data from sizing (peak height).

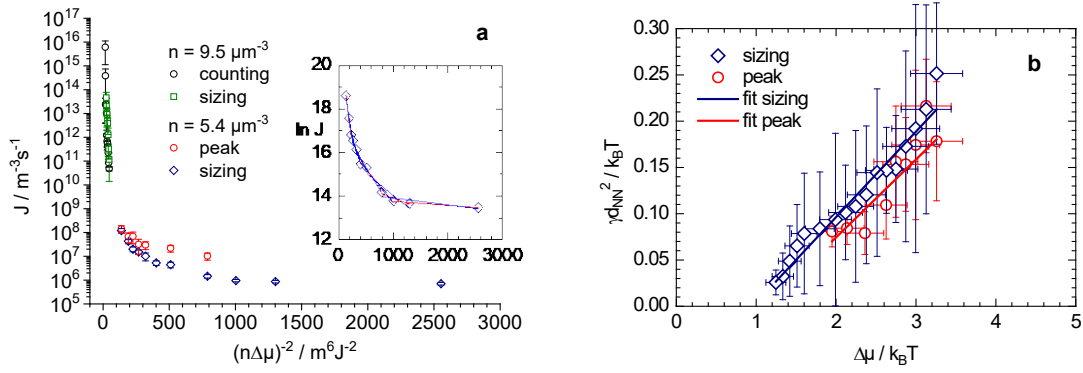

FIG. 5: CNT parameterization. Symbols as before and indicated in the key. (a) Nucleation rate densities plotted versus $1/(n\Delta\mu)^2$. Inset: natural logarithm of $J$ plotted versus $1/(n\Delta\mu)^2$ for the sizing data obtained at $n = 5.4\mu m^{-3}$. Solid lines are least squares fits through three (red) and four (blue) points. (b) Reduced IFE in units of the thermal energy as obtained from the fits versus metastability for the data taken at $n = 5.4\mu m^{-3}$. Solid lines are least squares linear fits returning the slope $m = C_T$.

Such a behaviour is at variance to previous findings on charged spheres. In Fig. 6(a), we compare it to two other systems, PnBAPS70 [40, 50, 75] and PnBAPS68 [29] both investigated in dependence on number density. There, the reduced IFE is much larger and extrapolates to a positive $\sigma_0$ at zero $\Delta\mu$. The here obtained values for $\sigma$ are even significantly smaller than the IFE of hard spheres (dashed line) and extrapolate to negative $\sigma$. Further, a Turnbull coefficient of $C_T \approx 0.1$ is much lower than the value of 0.3 observed before, which was in good agreement with that found for bcc crystallizing metals [76]. In Fig. 6(b) we plot the critical radii calculated as $r_{crit} = 2\gamma/n\Delta\mu$. For PS109*, this parameter increases with metastability, whereas for PnBAPS68 and PnBAPS70 it follows the CNT expectation for homogeneous bulk nucleation showing a decrease with $\Delta\mu$. Figure 6(c) shows curves of the free energy as a function of the nucleus radius $r$ calculated as: $\Delta G^* = -(4\pi/3)r^3 n\Delta\mu + 4\pi r^2 \gamma$. For deionised PnBAPS68 at $n = 60.9\mu m^{-3}$, $\Delta\mu \approx 9.9k_BT$ and $\gamma \approx 65k_BT\mu m^{-2}$, the initial rise is well pronounced. A maximum value of $\Delta G^* \approx 12.7k_BT$ is reached at $r_{crit} = 0.22\mu m$. Barriers at lower $\Delta\mu$ are even much larger for PNBAPS68. By contrast, for PS109* the curves are much shallower. Maxima are all well below $1k_BT$, indicating a practically barrierless nucleation process. In fact, they appear to vanish completely as the salt concentration is increased, as indicated in the inset by the arrow.

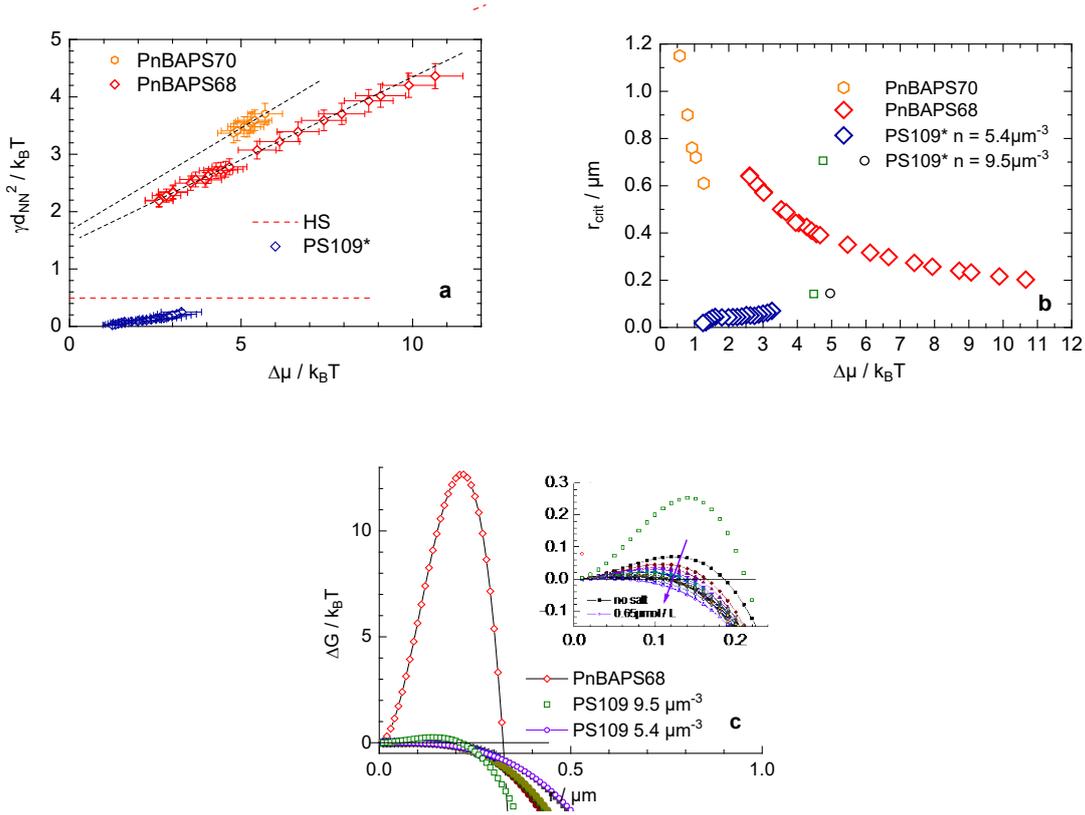

FIG. 6: Comparison of CNT parameters to those obtained for other charged sphere species. (a) Reduced IFE in dependence on Δμ. Data for PS109* taken at $n = 5.4\mu m^{-3}$ are compared to two data sets from previous measurements, in which good agreement with the CNT expectations for homogeneous bulk nucleation was observed: PnBAPS68 [29] and PnBAPS70 [75]. Both latter systems show much larger reduced IFE. Moreover, these extrapolate to a reduced equilibrium IFE of $1.5k_BT$ and $1.7 k_BT$, respectively. We also display the IFE of HS systems [4, 77] which amounts to $0.51k_BT$ (dashed red line). (b) Critical radii in dependence on Δμ. PnBAPS68 [29] and PnBPaPS70 [50, 75] illustrate the behaviour expected from CNT for homogeneous bulk nucleation as seen in many previous systems before: $r_{crit}$ decreases with Δμ. For PS109*, $r_{crit}$ increases with Δμ. (c) Nucleus free energies ΔG in dependence on nucleus radius as calculated from the obtained CNT parameters. Symbols denote: red squares: deionized PnBAPS68 at $n = 60.9$ $\mu m^{-3}$; open green squares: PS109* at $n = 9.5\mu m^{-3}$; open violet circles: PS109* at $n = 5.4\mu m^{-3}$ and $c_s = 0.65\mu molL^{-1}$. The inset shows a magnification of the free energy curves of PS109* at $n = 5.4\mu m^{-3}$ with the salt concentration increasing from zero (black solid squares, upper curve at $n = 5.4\mu m^{-3}$) to $c_s = 0.65\mu molL^{-1}$ (open violet squares, lowest curve). The arrow denotes increasing salt concentration. Note the low maximum values for PS109* irrespective of sample parameters and the tendency to decrease further with increasing salt concentration.

To summarize our observations for PS109*, we investigated a fully inconspicuous, highly charged system of comparably low polydispersity in the region of low melt metastability including parts of the coexistence region. Bulk nucleation rates determined from different approaches vary for increased salt concentrations from $J = 10^{16} m^{-3} s^{-1}$ to $J = 10^{11} m^{-3} s^{-1}$ at $n = 9.5 \mu m^{-3}$ and from $J = 10^{8} m^{-3} s^{-1}$ to $J = 10^{6} m^{-3} s^{-1}$ at $n = 5.4 \mu m^{-3}$. We observe peaked nucleation of bulk crystals and a crystallite size distribution skewed towards larger crystals way into the coexistence region. Parameterization along the lines of CNT yielded quantitative and qualitative deviations from previous experiments as well as from the CNT expectations for homogeneous nucleation. In fact, according to this parameterization, the nucleation events in our experiments appear to be practically barrierless. We conclude that crystallization in PS109* does not follow the standard path for bulk homogeneous nucleation. Based on these deviations and the finding of candidate seed particles, we tentatively suggest to further rationalize the present observations in terms of nucleation at particle doublets.

**D. Heterogeneous nucleation at doublets**

At the investigated particle densities, doublet sizes are smaller than the experimental interparticle distances. The seed size thus appears to be much too small to discuss our observations in terms of classical heterogeneous nucleation concepts affording the existence of an extended surface [7, 35]. In fact, such seeds would be much smaller than in any previously addressed case and, moreover, carry a larger charge than the embedding melt particles. We therefore next discuss their integration in the melt and crystal phase before proceeding with parameterization. In the present experiments, the question of integration is one of "solvation" rather than of wetting. Successful nucleation requires that the doublets are neatly integrated into and not expelled from both fluid and crystal phase. Unfortunately, the present particles and doublets are too small to directly visualize their integration by high resolution optical microscopy. We therefore resort to experiments, on binary CS systems of similar size ratio, $\Gamma = a_{small}/a_{large}$. Three instructive examples are shown in Fig. A3 (a) to (c) of Appendix B. They reveal excellent solvability of odd sized objects by low salt CS crystals and thus their principal ability to act as heterogeneous seeds [78, 79]. Homogeneous and heterogeneous doublets are observed to be neatly embedded into the lattice structure without visible distortion of the latter (Fig. A3(a) and (b)). Such a behaviour is known neither for atomic nor hard sphere systems. It is here enabled by fact that the observed CS crystallize already at very low volume fractions [12, 15]. Then, the lattice constants by far exceed the particle sizes and odd particles can be integrated without significant elastic energy penalty. Moreover, larger particles are frequently found throughout the complete crystals (Fig. A3(c)). This again demonstrates their excellent solubility in the surrounding crystal phase. Consistent with reports in the literature [80, 81], odd particles are also (partially) expelled into the grain boundaries, when their solubility limit is crossed (Fig. A3(c)). However, this occurs after nucleation during both the growth stage and the coarsening stage following complete solidification. It indicates, that the initially formed crystals containing the odd particles are obviously mechanically stable, but not so thermodynamically [82]. Finally, also much larger CS may act as heterogeneous nucleation seeds [40, 75].

Further evidence for good solubility of odd-sized particles in crystalline phases stems from bulk studies on the phase behaviour of size disperse CS systems and of binary CS mixtures. In fact, the terminal polydispersity for the thermodynamic stability of colloidal crystals is $PI_{term}$ ≤ 13% for CS as compared to some 7% for HS [83]. In a recent simulation study [23], an excellent solubility of smaller particles in CS bcc crystals was observed at dopant fractions $f_{small}$ ≤ 0.01. Only above, phase separation occurred. Further, we previously investigated the solidification of systems with added larger particles at $\Gamma = 0.57$. These were quenched into the liquid-solid coexistence region of the eutectic equilibrium phase diagram at $f_{large} \approx 0.05$ [56]. Only systems in flat cell confinement and under prolonged influence of gravity phase separated on long time scales of weeks to months, to form $bcc_{small}$ and $bcc_{large}$ phases coexisting with a fluid phase layered by gravity [84]. Bulk systems, however, quickly solidified completely to form substitutional alloy crystals of bcc structure. Together, these studies show the excellent solvability and integration of larger particles in a CS crystal lattice. They thus strongly support our suggestion that in the present experiments we observed doublet-induced nucleation.

A useful parameter to characterize seeded nucleation is the so-called seeding efficiency $\eta = \rho / n_{seed}$ given as the ratio of the past solidification crystallite density to the seed density. For PnBAPS109*, we estimate an upper bound for the seed fraction by equating it to the doublet fraction $f_D = f_{seed} \leq 10^{-3}$. The seed density then reads $n_{seed} = nf_{seed}$. We initially plotted η as a function of $\Delta\mu$ (c.f. Fig. A5 in Appendix D). That resulted in two data groups for the two $n$. Each increased exponentially with $\Delta\mu$ but the three orders of magnitude discontinuity remained at the same meta-stability. However, when plotting η in dependence on $n\Delta\mu$ in Figure 7(a), all seeding efficiencies arrange on a single curve (dashed line shown for comparison). Within experimental uncertainty, η increases exponentially with the product of metastability and number density for both experiments. All efficiencies, however, stay below unity and approach it from below.

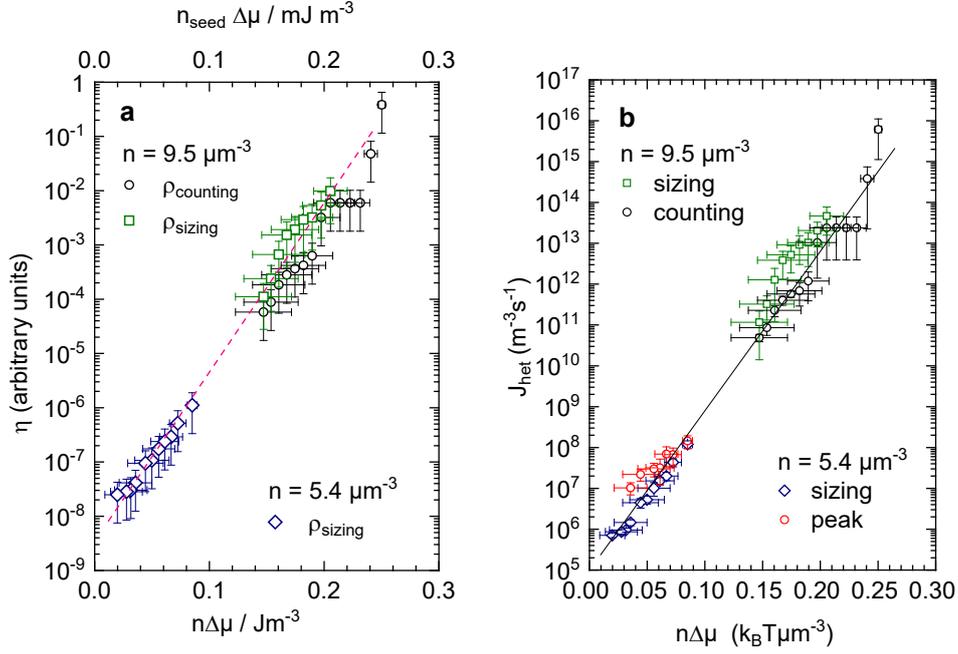

FIG. 7: Seeded nucleation: (a) Efficiency of seeded nucleation as a function of $n\Delta\mu$ (lower scale) respectively $n_{seed}\Delta\mu$ (upper scale). Data arrange on a straight line (dashed). With increasing $n\Delta\mu$, efficiencies approach unity from below. (b) Heterogeneous nucleation rate densities as a function of $n\Delta\mu$. Data arrange on a straight line (solid for comparison) indicating an exponential dependence of $J_{het}$ on $n\Delta\mu$.

A previous study [40] investigated deionized systems of charged spheres (PnBAPS70, $Z_{eff}$ = 450e$^-$) at large $n$ = 22.4µm$^{-3}$, implying large metastability. Being way above the coexistence density $n_F$ = 5µm$^{-3}$, this implied a zero density difference between melt and solid. At constant $n$, the authors added various small amounts of large ($2a$ = 15µm), completely wetted seeds ($f_{seed} = 10^{-11} - 10^{-7}$). For increasing seed concentrations, they reported a superseeding of homogeneous nucleation by heterogeneous nucleation. Simultaneously, the character of nucleation kinetics switched from steady state to peaked. Peak rates increased stronger than exponentially with $n_{seed}$ in the transition region, indicating an increase in efficiency. Above the transition, their crystallite density scaled linearly with their seed density. There in fact, η was observed to be very close to unity, i.e. all seeds successfully nucleated a crystal [40]. Our study complements this work. We worked at low metastabilities and explicitly cover nucleation in the coexistence region. The present suspensions contained much larger seed fractions but, due to low metastability, the nucleation rates stayed rather small. In addition, across coexistence, our time dependent nucleation rates were quenched already at an early stage. That was attributed to the small but significant density difference between melt and crystal phase across coexistence related to an adjustment of the melt density in the course of solidification. Thus, despite good compatibility with mother and daughter phase, not all doublets could actually form a crystal during the initial stage up to the fast quench in nucleation rate. In fact, the majority of our crystal phase formed during the growth stage. From both studies, it appears, that the factors positively influencing the efficiency are small seed fractions, significant metastability, absence of density differences and nucleation dominated solidification through large nucleation rates.

Next, we plot in Fig. 7(b) the heterogeneous nucleation rate density, $J_{het}$, versus $n\Delta\mu$. Also here the data show a single exponential increase with the product of number density and metastability. Within experimental uncertainty, we find the nucleation rates to be compatible with the empirical relation: $J_{het} \propto \exp(n\Delta\mu)$. While the dependence on the chemical potential difference is expected, the dependence on $n$ is surprising. For constant efficiency, one would have the trivial linear dependence due to $n_{seed} = n f_{seed}$. However, we are not aware that an additional exponential dependence of $J_{het}$ on $n$ has been observed or predicted before. We thus can only speculate about its origin.

We recall that the equilibrium structure of CS fluids and melts is bcc-like [85, 86], i.e. already close to that of the target phase. This structure gets more pronounced upon reducing the salt concentration but also more extended upon increasing the number density. Melt pre-structuring in CS has been discussed before, e.g. to explain the observed preference for the body centred cubic (bcc) polymorph formation via homogeneous nucleation where fcc is the stable crystal phase [87]. Melt pre-structuring is also an issue in the homogeneous nucleation of HS [20] and nucleus growth control in silicon [24]. With more pronounced local structure, also the collective dynamics of the melt slow down [88]. In fact, structure and dynamics are closely coupled by scaling relations and the onset of crystallization mirrored in a dynamic freezing criterion equivalent to the Hansen-Verlet criterion [74]. Incidentally, the former was originally observed in PnBAPS109* [53], but since has been demonstrated also in many other systems including HS and dusty plasmas [89, 90, 91]. Local slowing of dynamics has been explicitly related to nucleus formation in a recent simulation work on ice nucleation in water [92]. There, the authors observed that ice nucleation occurred in low-mobility regions distrib-uted throughout the liquid. Further, there is a dynamical incubation period in which the mobility of the molecules drops before any ice-like ordering. And finally, ice-like clusters, once formed, cause arrested dynamics in surrounding water molecules. The authors further suggested that similar mechanisms could also be at work in water containing dissolved molecules. Based on these results from literature, we anticipate that also a doublet, still small, but of higher than average charge, can increase the local structure, slow down local dynamics and favour nucleus formation. We expect the seed-induced enhancement of local structure to be itself dependent on the pre-ordering of the melt. The kinetics for this type of nucleation process would then depend on both the chemical potential difference $\Delta\mu$ and the $n$-dependent degree of additional melt pre-structuring by the doublet. The former would as usual control the basic nucleation rate. The latter would control the increased probability of heterogeneous nucleation in regions pre-structured by the seeds on top of homogeneous nucleation. This speculation clearly leaves the trodden path of CNT, but may inspire tests by extended as well as new experiments and theoretical work.

Future studies should first of all address further salt concentration dependent nucleation experiments at low meta-stabilities. Here doublet free systems conditioned in a circuit equipped with online filtering should be addressed, as well as systems with seeds of like size but larger charge than those of the host suspension. Second, employing systems with larger particles, experiments using confocal (fluorescence) microscopy should reveal valuable information about the local structure of the melt close to a seed and facilitate comparison to the overall structure at metastable conditions. Local dynamics and possible dynamical heterogeneities could be investigated using dynamical differential microscopy analysis. Thirdly, it remains to be explored in experiments at moderate to large metastability, whether the exponential dependence of $\eta$ and $J_{het}$ on $n$ is retained throughout, or whether at some point it switches to constancy for $\eta$ and a linear dependency for $J_{het}$ through $n_{seed} = n f_{seed}$.

The important role of simulations to study colloidal crystallization in parallel to optical experiments has been mentioned [19, 20, 21, 22, 23, 24, 25, 26, 92]. Concerning metastable CS melts and their microscopic crystallization kinetics, it appears to be rewarding to study the changes in local structure and dynamics induced by individual particles of like size but larger charge and/or of particle doublets. This could be used to test our speculative suggestions in a more general way.

Finally, our observations may also bear some relevance to nucleation in strongly polydisperse samples. Also here, larger and more highly charged particles are present, which may in principle act as "self-heterogeneous" seeds through a mechanism similar to that suggested above. If this applies, one can expect an initial accelerated increase of $J$ with $\Delta\mu$ followed by a transition to a nearly constant rate density. This could be tested either by simulation or in nucleation experiments combining microscopy, time-resolved static light scattering, and reflection microscopy to cover an extended range of number densities.

## IV. CONCLUSIONS

The present study pioneered salt concentration dependent measurements of colloidal crystal nucleation rate densities at low metastabilities. We employed a continuous conditioning under conductometric control to vary the concentration of added electrolyte at constant particle concentration. This incidentally introduced a small fraction of doublets. We further combined time resolved Bragg microscopy, post solidification size analysis and crystallite counting to characterize the solidification kinetics. Data from these different approaches gave consistent results. We observed peaked nucleation to prevail at all conditions. Analysis in terms of homogeneous CNT yielded unusually low nucleus-melt interfacial free energies and a vanishing nucleation barrier. The nucleation behaviour thus differed qualitatively from that known for homogeneous nucleation of similar systems. We attribute our findings to nucleation at doublets. Data on their seeding efficiency are complementing previous findings for the seeding efficiency of large spherical seeds at larger metastabilities. In the present study the efficiency stayed below unity and we obtained the empirical relation, $J_{het} \propto \exp(n\Delta\mu)$. On the basis of our results, the possibility that local, seed-induced structural enhancement and dynamic slowing together lead to this exponential increase in nucleation rates was cautiously suggested. Additional systematic experiments on the kinetics of seed induced nucleation as well as studies on local melt structure and dynamics are desired to test this speculative idea. We anticipate that our study may also stimulate theoretical interest. Employing recently published approaches to investigate in general the triggering of bulk nucleation by very small objects seems to highly rewarding.

ACKNOWLEDGEMENTS

We thank L. Shapran for contributing to the effective charge characterization, H. J. Schöpe for assistance in data evaluation and M. Evers for the composition analysis. P. Wette and D. Herlach gave valuable input in experimental issues. We thank R. Biehl, A. Engelbrecht, H-Reiber, and A. Reinmüller for kindly providing image material from their PhD theses. We are further indebted to our theoretical colleagues, with whom we had numerous interesting discussions: J. Horbach, H. Löwen, T. Schilling. Financial support of the DFG is gratefully acknowledged.

**Conflict of Interests**

The authors declare no conflict of interest.

**Data Availability Statement**

Original data are available from the corresponding author upon reasonable request.

# APPENDIX A: ADDITIONAL DETAILS OF SYSTEM CONDITIONING AND CHARACTERIZATION

The original sample was shipped at a packing fraction of $\Phi \approx 0.08$. By dilution with distilled water, we prepared suspensions of approximately $\Phi = 0.01$, added mixed-bed ion exchange resin (IEX) [Amberlite UP 604, Rohm & Haas, France] and left it to stand with occasional stirring for some weeks. The suspension then was filtered using Millipore 0.5 μm filters to remove dust, ion-exchange debris and coagulate regularly occurring upon first contact of suspension with IEX. A second batch of carefully cleaned IEX filled into a dialysis bag was then added to retain low ionic strength in the stock suspensions now kept under Ar atmosphere.

Before use, samples are coarsely filtered upon filling the circuit to remove dust and coagulate and other impurities remaining from pre-conditioning. All further sample preparation and the measurements were performed in a closed system including the measuring cells and the preparation units [60, 93]. A schematic drawing for the arrangement used for the solidification experiments is given in Fig. A1(a). The suspension is pumped by a peristaltic pump (P) through a closed and gas tight Teflon® tubing system (black arrows) connecting i) the ion exchange chamber (IEX); ii) a reservoir under inert gas atmosphere to add suspension, water or salt solutions (R); iii) a cell for conductivity measurements (C) and iv) one or more cells for the optical experiments ($OC_i$). All components are equipped with gas tight tube fittings, either custom made or commercial (Bohländer, Germany). By default, the circuit is further equipped with an online filter integrated past the pump (not shown). Pore size is chosen roughly three to four times the particle diameter, to avoid extensive clogging. Then, the exact number density is monitored online *via* an integrated static light scattering experiment (e.g. $OC_1$). For the nucleation and growth experiments, this filter component was omitted to avoid particle loss during conditioning and to work at strictly constant number density. Then, particle aggregates and ion exchange resin splinters are not removed, but accumulate with time. For larger particles, this can be directly seen in microscopy. An example is shown in Fig. A1(b). Under deionized or low salt conditions, the majority of aggregates are doublets. For PnBAPS109* we assessed their number fraction as a function of time by AFM analysis as described in the main text.

Typically, the suspension is cycled until a stable minimum conductivity is reached, defining the thoroughly deionized state. Then the ion exchange chamber is bypassed, using the three way valves (V). If desired, salt solution can be added to adjust the electrolyte concentration under conductometric control. Care is taken to assure a $CO_2$-leak free circuit. Circuit preparation facilitates stable experimental conditions on a typical time scale of an hour, during which different measurements can be performed. In the salt concentration dependent measurements of the present work, we deionized the suspension again after each crystallization run. This avoids dilution errors in the salt concentration and accumulation of leaked carbonic acid.

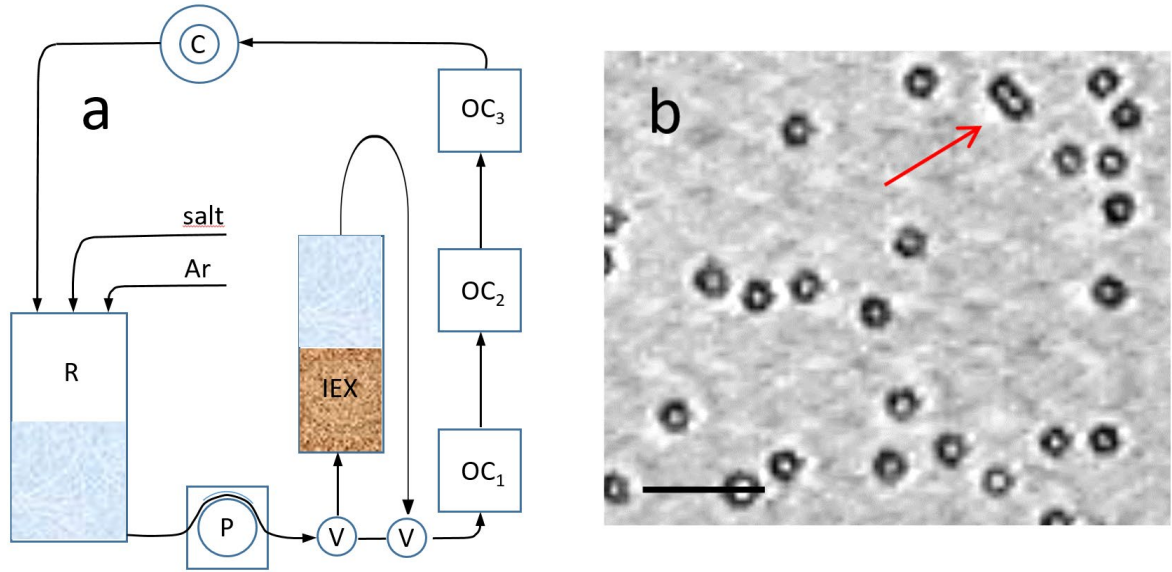

FIG A1. (a) Conditioning circuit as used in the present crystallization experiments. R: reservoir, P: peristaltic pump, IEX: ion exchange column, C: conductometric cell, $OC_1$, $OC_2$, $OC_3$, … optical flow through cells. All components are connected by gas tight tubings, with flow direction indicated as arrows. After thorough deionization the IEX can be bypassed using the three way valves (V). Then, salt solution can be added to the reservoir under an inert gas atmosphere and its concentration monitored *via* conductivity. (b) Typical result of prolonged cycling in the peristaltically driven conditioning circuit without integrated filter. The image shows a cropped bright field transmission image (Leitz, N PLAN 10×/0.25; standard video camera). The scale bar is 2.5µm. The arrow highlights a doublet formed after some cycling in a dilute suspension ($\Phi = 10^{-4}$) of PnBAPS359 (BASF, Ludwigshafen, Germany).

For static and dynamic light scattering, as well as for Torsional Resonance Spectroscopy [94], we used quartz cells of 1cm outer diameter mounted to a multipurpose light scattering set-up described elsewhere [95]. This instrument allows measurements of the form factor of non-interacting suspensions, of the structure factor of fluid samples, and – for crystalline samples – of the particle number density from the position of Bragg reflections as well as of the shear modulus from the location of mechanical resonances in frequency space.

Conductivity is measured at a frequency of ω = 400 Hz [electrodes LTA01 and LR325/01 with bridge LF538 or electrode LR325/001 with bridge LF340, WTW, Germany]. To check for reproducibility we compared conductivity values at different frequencies ω ≤ 1 kHz but found no dependence on ω. Thus ω is low enough to measure the DC-limit of conductivity but large enough to inhibit significant electrophoretic particle transport. Care was further taken to control the suspension temperature within 1K. In general the reproducibility of conductivity measurements in suspensions was found to be better than 2%. Fig. A2(a) shows the conductivity of PS109* in the deionized state as a function of increased number density. We obtain a linear relation with an offset of $\sigma_B \approx 60$ nScm$^{-1}$ due to water hydrolysis. In the deionized state, the $n$-dependent conductivity reads [57, 58]:

$$\sigma = \sigma_0 + \sigma_B = neZ_\sigma(\mu_P + \mu_{H+}) + \sigma_B \quad (A1);$$

where $\mu_P$ and $\mu_{H+}$ are the electro-phoretic particle and proton mobility, respectively. The least squares fit to the data returns $Z_\sigma = 459 \pm 20$ e$^-$. When, after deionization, the electrolyte concentration is adjusted by adding small amounts of NaCl solution, the exact concentration is inferred from the increase in conductivity using again Hessinger's model of independent ion migration.

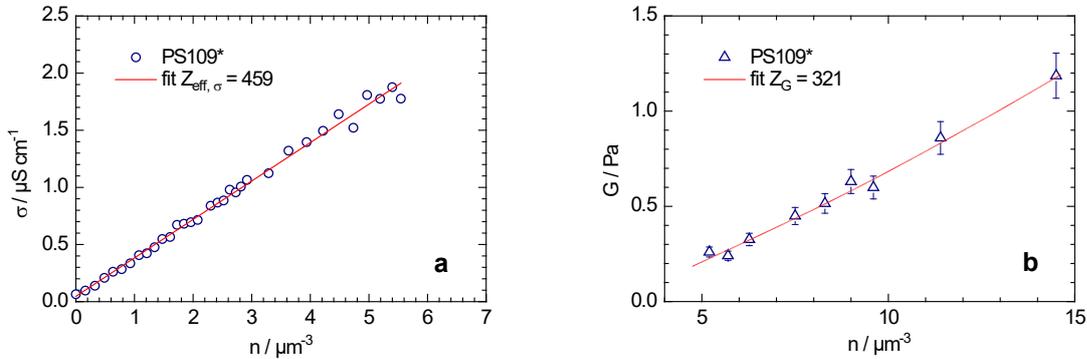

FIG A2: system characterization. (a) Sample conductivity as a function of number density. The red solid line is a least square fit of Eqn. (A1) to the data returning $Z_\sigma = 459 \pm 20$ e$^-$. (b) Sample shear modulus as a function of number density. The red solid line is a least squares fit of the Eqn. (A5) to the data returning $Z_G = 321 \pm 18$ e$^-$.

The pair energy of interaction is modelled as a screened electrostatic repulsion neglecting van der Waals attractions. We use a mean field expression derived from the Debye-Hückel solution for ions of finite radius:

$$V(r) = \frac{Z_G^2 e^2}{4\pi\varepsilon} \left(\frac{\exp(\kappa a)}{1+\kappa a}\right)^2 \frac{\exp(-\kappa r)}{r} \tag{A2}$$

Here, e is the elementary charge, and $\varepsilon = \varepsilon_0 \varepsilon_r$ is the solvent dielectric permittivity. The screening parameter $\kappa$ is defined as:

$$\kappa = \frac{e^2}{\varepsilon k_B T} \sqrt{n Z_\sigma z^2 + n_S z^2} \tag{A3},$$

where $z = 1$ is the micro-ion valency, $n$ is the particle number density. The micro-ion number density, $n_s$, is calculated accounting for ions stemming from added electrolyte and dissolved $CO_2$ as well as ions from the self dissociation of the solvent. Note that the counter-ion density, which is explicitly accounted for through the $nZ_\sigma$ term, contributes the majority of screening ions in our experiment.

This description assumes that individual ionic clouds overlap without distortion. Many-body effects arising from Double layer overlap are, however, accounted for by using $Z_G$ instead of $Z_\sigma$ in Eqn. (A3). The former is determined in crystalline samples using Torsional Resonance Spectroscopy. The cylindrical sample cell containing a polycrystalline colloidal solid is put into low amplitude oscillations about its vertical axis. As the shear moduli of the sample are small, standing waves can be excited with wave lengths comparable to the container dimensions [51]. On the colloidal level, this corresponds to harmonic lattice vibrations visible as shifts of the Bragg-peaks or distortions of the first structure factor peak. These are detected and frequency analysed by time-resolved static light scattering in combination with lock-in technique. Resonance frequencies are observed at:

$$\omega_{jm}^2 = \frac{G(\mu_j^2 + (m+1)^2 \pi^2 \alpha^2)}{\rho R^2} \tag{A4},$$

where $j$ and $m$ are the indices of the order of the resonances, $\alpha = 0.5$ is the ratio between the cell radius $R$ and the filling height $H$, $\rho$ is the mass density of the suspension, and the $\mu_j$ denote the zeros of the 1st order Bessel-function $J_1$.

Figure A1(b) shows the shear modulus $G$ of deionised suspensions of PS109* as a function of $n$ as determined from static light scattering. It increases from $G = 0.3$Pa at n = 5µm$^{-3}$ to $G =$

1.2Pa at n = 14.5µm⁻³. Each point represents an average over 3 to 5 measurements, the residual errors are about 5 %. The shear modulus $G$ for crystals of bcc symmetry reads:

$$G_{bcc} = f_A \frac{4}{9} nV(d_{bcc}) \kappa^2 d_{bcc}^2 \tag{A5}$$

$f_A = 0.5$.is a numerical factor which accounts for the different boundary conditions in averaging over randomly oriented crystallites or local environs, and the nearest neighbour distance in a bcc lattice reads:

$$d_{bcc} = \frac{\sqrt{3}}{\sqrt[3]{4\,n}} \tag{A6}$$

Equation (A5) is then used in a least squares fit to the data returning $Z_G = 321 \pm 18 e^-$.

## APPENDIX B: COLLOIDAL SUSPENSIONS CONTAINING LARGER OBJECTS

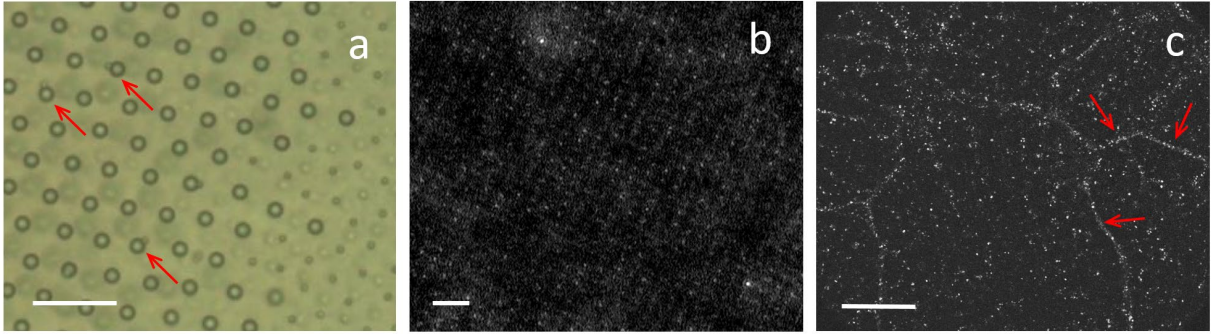

Fig. A3: Larger Objects in colloidal crystals and fluids as observed by different microscopic techniques. (a) Heterogenous doublets. (b) homogeneous doublets from conditioning. (c) Larger spheres.

Fig. A3(a) displays a 1:1 mixture of thoroughly deionized PS2.9µm und PS1.1µm (microparticles Berlin GmbH, Berlin Germany). We show a cropped bright field transmission image (Leitz, FLUOTAR L 63x/0,7; Nikon D800E) The scale bar is 10µm. This two layer system shows macroscopic phase separation [78]. To the right, there is a fluid of PS1.1µm particles. To the left we see the lower layer of a wall based fcc crystal of PS2.9µm particles. Its adjacent layer is faintly visible. Both regions are separated by a rough interface. In addition to the pure components, one further sees a number of mixed doublets (arrows). These are all well integrated into the PS2.9μm lattice, indicating their excellent wettability.

Fig. A3(b) displays a bcc wall crystal of thoroughly deionized PS301 (IDC, Portland, OR). We show a dark field transmission image (Leitz, PL Fluotar L 63x/0.7 corr PH2 1/0.1–1.3/C): The scale bar is 10μm. The image was taken with an enhanced depth of field [79]. It shows a few layers of the (110) lattice planes of a wall based bcc crystal grown parallel to the lower substrate. At the upper right and the lower left, two conditioning related doublets visible as bright spots. Note, that the crystal structure is not distorted by the presence of the doublets. This shows that the open bcc structure of lattice constant 2.8μm ≈ $20a$ readily accomodates the larger objects during the crystal growth stage. Again this indicates complete wetting.

Fig. A3(c) displays a polycrystalline suspension formed in a deionized 20:1 mixture of PnBAPS70 (BASF, Ludwigshafen, Germany) and Europium-Chelate-COOH-dyed PS200 (FCEU002, Bangs-Lab Dr. Fishers). We show a cropped fluorescence microscopic image (Leitz, HC PL FLUOTAR 10x/0,32 PH1, $\lambda_{cutoff}$ = 580nm; Nikon D800E). The scale bar is 250μm. This eutectic system shows phase separation on long time scales (weeks). However, it initially solidifies as substitutional alloy crystal of bcc structure, when quenched into the equilibrium fluid-crystal state. The formed crystals are therefore metastable. The image shown was taken about 30 min after complete solidification. The larger, fluorescently dyed particles are visible as bright spots. At this size ratio of $\Gamma \approx 0.3$ and lattice constants of $\approx 5a$, particles are found both throughout the crystal grains (arrows) and in the grain boundaries. On the timescale of days, more and more fluorescent particles accumulate in the grain boundaries. Similar effects have also been observed in other systems. For instance, Ghofraniha et al [80] investigated grain refinement in crystallized 22nm charged micelle systems with 1%-2% of added 19% polydisperse silica particles of 30nm average size. On the time scale of days, the crystallite size increased, while most of the silica particles became segregated into the grain boundaries. Thus again, crystallites were born contaminated. In selected cases, the authors could identify a central particle using high resolution DIC microscopy [81], indicating their wettability. While soft micellar systems are not directly comparable to CS, their studies nevertheless clearly evidence the possibility of heterogeneous nucleation by similarly sized objects. Examples of wetting and heterogeneous nucleation at much larger CS seeds can be found in [40, 75]. Together these examples demonstrate that CS suspensions readily wet individual odd sized objects irrespective of their structure. The objects in turn can then act as heterogeneous seed particles. They further can become embedded during growth or expelled to grain boundaries during growth and ripening. However, such a precipitation into grain boundaries or a second phase occurs at solidification stages later than nucleation.

APPENDIC C: CNT FOR COLLOIDAL SYSTEMS

CNT originally was originally designed to model condensation from the vapour phase. It was first adapted to the case of crystallization [67,68,69,70, 71,72] in atomic systems and later also to colloidal crystallization respecting their specific interactions and their diffusive dynamics [32]. Russel [96] approximated the kinetic pre-factor by an attempt frequency for diffusional relaxation: $J_0 = nD_{eff}/\ell^2$, where $D_{eff}$ is an appropriate diffusion coefficient and $\ell$ the corresponding length scale. Schöpe and Wette [29] further equated $\ell^2$ to $n^{-2/3}$ and considered the barrier shape *via* the Zeldovich factor:

$$J_0 = 1.55 \cdot 10^{11} \cdot n^{4/3} \sqrt{\gamma} \cdot D_S^L(n) \tag{A7}.$$

The CNT expression for heterogeneous nucleation at extended surfaces reads [7]:

$$J = J_0 f_{seed} \frac{\alpha(\phi)}{\sqrt{f(\phi)}} \exp\left(-\frac{\Delta G^*}{k_B T} f(\phi)\right), \tag{A8},$$

where $\phi$ is the contact angle of the crystal phase on the substrate, and $f(\phi)$ varies from 0 for $\phi = 0$ to 1 for $\phi = \pi$. The function $\alpha(\phi)$ refers to the surface area of the daughter phase as a function of $\phi$. The special case of complete wetting on spherical substrates was first worked out in detail by Fletcher [35] and recently re-examined in [97]. Here, $f(\phi,X)$ replaces $f(\phi)$, with $X = R_{seed}/r$ being the ratio of seed size to nucleus size, and $\alpha(\phi,X)/(f(\phi,X))^{1/2}$ is a factor of order unity. Note, that in that case, the work of formation of the critical nucleus is defined only if $R_{seed} > r_{crit}$. As $X$ approaches unity from below, i.e. upon adsorption of a first layer of crystal phase $f(\phi)$ drops steeply towards zero. Hence, the nucleation barrier in Eqn. (A8) vanishes. The case of very small seeds has not yet been considered in CNT.

APPENDIX D: ADDITIONAL DATA FROM CNT PARAMETERIZATION

This section contains additional data and intermediate results obtained during parameterization of both nucleation experiments with CNT.

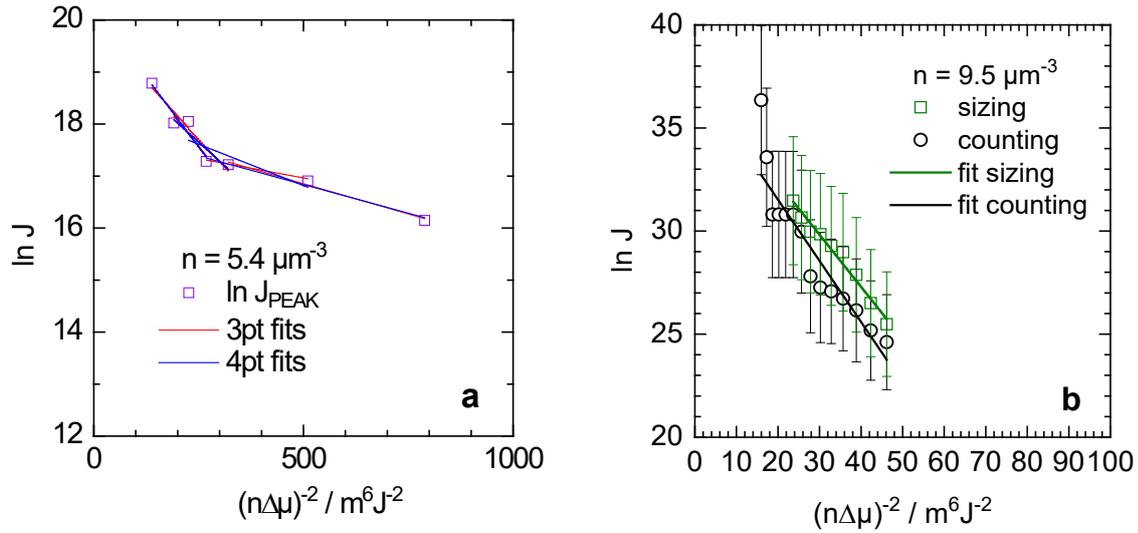

Fig. A4: CNT parameterization. (a) Natural logarithm of $J$ plotted versus $1/(n\Delta\mu)^2$ for the peak data obtained at $n = 5.4\,\mu m^{-3}$. Solid lines are least squares fits through three (red) and four (blue) points. (b) Natural logarithm of $J$ plotted versus $1/(n\Delta\mu)^2$ for the sizing (olive) and counting (black) data taken at $n = 9.5\,\mu m^{-3}$. Solid lines are least squares fits.

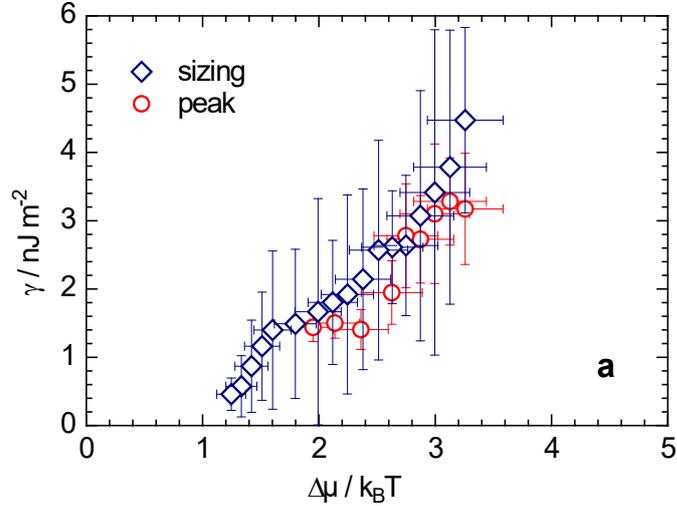

Fig. A5: Interfacial free energy as a function of metastability for the sizing and peak data obtained at $n = 5.4\,\mu m^{-3}$ as denoted by the colour code indicated in the key.

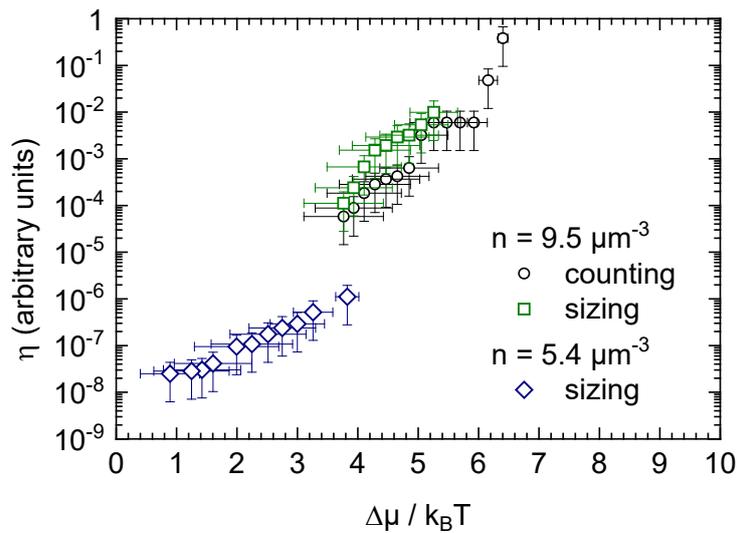

Fig. A6: Seeding efficiency calculated assuming a constant seed fraction in dependence on metastability. Data shown are for two different number densities and determined by different evaluation approaches, as indicated in the key. Note the pronounced discontinuity at $\Delta\mu = 4k_BT$.